\documentclass{statsoc}

\usepackage[a4paper]{geometry}
\usepackage{graphicx}
\usepackage[textwidth=8em,textsize=small]{todonotes}
\usepackage{amsmath,bm}
\usepackage{amssymb,amsfonts,bbm}
\usepackage{natbib}
\usepackage{setspace,url}
\usepackage{verbatim,mathrsfs,float,epstopdf,moreverb,multicol}
\usepackage{lscape}
\usepackage{xcolor}

\newcommand{\be}{\begin{enumerate}}
\newcommand{\ee}{\end{enumerate}}

\title[Probabilities of historical migrations]{Estimating conditional probabilities of historical migrations in the transatlantic slave trade using kriging and Markov decision process models}
\author{Ashton Wiens}
\address{Department of Applied Mathematics, University of Colorado Boulder,
Boulder, Colorado,
USA.}
\author{Henry B. Lovejoy}
\address{Department of History, University of Colorado Boulder,
Boulder, Colorado,
USA.}
\author{Zachary Mullen}
\address{Department of Computer Science, University of Colorado Boulder,
Boulder, Colorado,
USA.}
\author[Wiens, Lovejoy, Mullen, and Vance] {Eric A. Vance}
\address{Laboratory for Interdisciplinary Statistical Analysis (LISA), \\
331 ECOT, 1111 Engineering Drive Box 526 \\
University of Colorado Boulder, Boulder, CO USA 90309-0526}
\email{Eric.Vance@Colorado.EDU}

\begin{document}
\begin{abstract}
Intra-African conflicts during the collapse of the kingdom of Oyo from 1817 to 1836 resulted in the enslavement of an estimated 121,000 people who were then transported to coastal ports via complex trade networks and loaded onto slave ships destined for the Americas. Historians have a good record of where these people went across the Atlantic, but little is known about where individuals were from or enslaved \textit{within} Africa. In this work, we develop a novel two-step statistical approach to describe the enslavement of people given documented violent conflict, the transport of enslaved peoples from their location of capture to their port of departure, and---given an enslaved individual's location of departure---that person's probability of origin. We combine spatial prediction of conflict density via Kriging with a Markov decision process characterising intra-African transportation. The results of this model can be visualised using an interactive web application, plotting estimated conditional probabilities of historical migrations during the African diaspora. These results help trace the uncertain origins of people enslaved in this region of Africa during this time period: using the two-step statistical methodology developed here provides a probabilistic answer to this question.
\end{abstract}
\keywords{Digital humanities; Gaussian process; Kernel density estimation; Markov decision process; Kingdom of Oyo; Transatlantic slave trade}

\section{Introduction}
The transatlantic slave trade consisted of over 12.5 million people boarding slave ships along the coast of Africa between 1514 and 1866 \citep{eltis2020,manning2020}. One of the egregious after-effects was the erasure of the identities of millions of Africans; children, women and men removed from their homes, forced in slavery, their names changed, their birth places and family ties obliterated. African historians have had great trouble understanding the origins of enslaved African people during the entire era of the transatlantic slave trade because of the scarcity of primary sources in the pre-colonial period \citep{lovejoy2011}. The size of inland populations during this period and the inland migrations that took place are unknown. This paper explores the internal African origins of enslaved people by focusing on a single quadrant of West Africa during the collapse of the kingdom called Oyo. This former slave trading state was located in what is now the Yoruba-speaking areas of southwestern Nigeria, as well as parts of Benin and Togo. According to Eltis in \textit{Voyages: The Trans-Atlantic Slave Trade Database} \citeyearpar{eltis2020}, an estimated 121,000 individuals from this region and time period involuntarily boarded slave ships, which primarily went to three destinations: Brazil (42\%), Cuba (40\%),  the British and French Caribbean (\textless 1 \%), and due to British abolition efforts after 1807, Freetown, Sierra Leone (17\%).

We have chosen to use Oyo as a test case for our methods to determine probabilities of origins of enslaved people because, due to early diplomatic and missionary activity, Oyo's collapse is relatively well-documented compared with other places and periods in Africa. Transient European explorers witnessed and documented warfare in Oyo, though without fully understanding the local or regional context \citep{clapperton2005,lander1830,lander1832}. Missionaries recorded narratives of enslaved Africans and oral accounts, but usually several year after they boarded slave ships and were liberated in British abolition efforts in Sierra Leone \citep{irving1856,johnson1921,ajayi1967,curtin1967,lloyd1967}. As Catherine Cameron has argued based on archaeological and bio-archaeological evidence, the enslavement of people around the world usually began following a violent act, most especially through warfare, raiding, or kidnapping \citep{cameron}.

Recently, Henry B. Lovejoy published an overview of the historiography and digital humanities methodologies used to analyze primary and secondary sources to extract geo-referenceable data of instances of conflict surrounding the turbulent era of Oyo's collapse \citeyearpar{lovejoy2019}. These annual maps of conflict illustrate the highest intensities of places of enslavement stemming from warfare in the Bight of Benin hinterland between 1817 and 1836. This recent breakthrough in compiling and mapping of intra-African conflict has unlocked the possibility of A) modelling how conflict can generate enslaved individuals who are B) transported through trade routes to ports of sale. We then use this model to C) determine likely origins of Africans transported involuntarily to the Americas or elsewhere within Africa on an annual basis. Inferring inland population movements will have a direct impact on linking together open-source collections of data, whether derived from historical sources, such at the \textit{Voyages} database, among others listed in Section \ref{S:Discussion}; or indeed, recently published results of genetic analyses of the African diaspora in the Americas \citep{23andme}.

Detailed shipping records of the transatlantic slave trade provide an annual quantitative baseline which sketches when and where people boarded slave ships at the coast and when and where they went in diaspora \citep{eltis2020,manning2020}. During Oyo's collapse, there are records for 243 voyages leaving ports in the Bight of Benin hinterland and amounting to documentation for approximately 75,700 people out of an estimated 121,000 people departing the coast between 1817 and 1836. These data allow for an excellent understanding of when and where more than half of the enslaved people from Oyo went by way of the Atlantic, but they contain many gaps and inconsistencies \citep{manning2015}, and they tell none of the story for the inland points of origin of these people. In addition, genetic studies have observed a concordance between ancestry based on genetic composition and the existing historical documentation of slave ship voyages \citep{23andme}. However, this inference is done at large spatial scales, while the work developed here is applied at a smaller scale making predictions about origin probabilities more precise. Furthermore, there is a gap in understanding the dynamics of inland population movements during this period, and we address this need using statistical modelling in this work.

Our approach unravels internal population movements through an analysis of conflict alongside documented slave ship departures annually to estimate the probability of inland origins of enslaved people on a port-by-port basis. We assume how instances of conflict at specific places would have generated enslaved people within Oyo territory, who then would have been transported to the coast via different trade routes.
As an analogy, our methodology operates like a watershed model in geology that predicts where precipitation collects and drains into a body of water \citep{nelson1994}; or more like catchment areas in human geography that describe the geographical boundaries of populations comprising, as an example, medical practices \citep{jenkins1996}. 
Our model describes where individuals may have been captured and enslaved in conflict in a given time period, and then predicts the port of departure accounting for the possible routes that could be taken to ports of departure. The focus of this paper is only on inland migrations to the coast, although this model could be applied to understand internal population movements into newly founded cities or elsewhere within Africa.

Knowing more about internal population movements and moments of conflict leading to enslavement could inform Africans and their descendants about their ancestry and heritage. This innovative methodology also gives a preliminary understanding of the possible scale and scope of internal population movements in this region of Africa where there is little documentation.

This novel approach reconstructs the internal slave trade during Oyo's collapse by synthesising spatial statistical models from points of historical conflicts and conjoining them with models for decision processes governing the inland transit of enslaved individuals. Our approach answers three questions relevant for historians and researchers of contemporary or historical forced migrations:

\begin{enumerate}
\item How can one model the enslavement of people or the creation of refugees given historical documentation of war and violent conflict? 

\item How can one model the transport of enslaved individuals, refugees, or migrants from their location of origin to their location of departure from the region? Or more generally, how can one model the migration of people using a sequential decision process?

\item Given an enslaved individual, refugee, or migrant's location of departure from a region, the conflicts in that region, and the possible transportation routes, what is that person's likely origin?
\end{enumerate}

In Section \ref{S:Data} we describe the historical data on conflict and trade routes in Oyo from 1817--1836. In Section \ref{S:Models} we describe the methods we use for creating conflict density functions, simulating enslavement from these functions, modelling the transit of enslaved individuals to points of sale, and creating maps of the conditional probabilities of origin given the point-of-sale. We describe our interactive web application for exploring the data and models in Section \ref{S:App} and present some illustrative results in Section \ref{S:Results}. Finally, we discuss in Section \ref{S:Discussion} how the models could be applied by historians, their relevance for other applications in the study of forced migrations, and potential extensions and improvements. Section \ref{S:Conclusion} concludes this paper.

\section{Historical Conflict, Trade Routes, and Slave Ship Data}
\label{S:Data}
This section describes several geopolitical data sets used in the work curated by historian Henry B. Lovejoy describing the enslavement, conflicts, and trade routes that existed during Oyo's demise from 1817--1836. First, we have annually recorded conflicts during the period, which may last one or more years. We have a trade network comprised of over 250 populated places which characterises the movement of people in the region, and this network is assumed valid during the entire period. We also have estimates of the total number of enslaved people departing the region as a whole and departing from specific trading ports, some with references to the years the journey took place. Finally, the visualizations produced in this work utilise shapefile data of prominent geographical features that existed in the region during 1817--1836. In particular, we include bodies of water in maps which are relevant to identifying the boundaries of the various states. These data were downloaded from http://www.diva-gis.org/ \citep{hijmans2001}. Several bodies of water created by damming long after the historical period under analysis were removed from the data set.

\subsection{Historical conflict in Oyo, 1817--1836}
\label{S:Data:Conflict}
Leading into the nineteenth century, Oyo was a major West African slave trading state that supplied tens of thousands of enslaved Africans to European slave traders at the coast. Prior to 1817, the kingdom had remained relatively stable. In 1817 this stability was disrupted by a Muslim slave uprising at Ilorin, which was the base of Oyo's major provincial army largely composed of enslaved Hausa from the north. Following a series of internal crises, jihad, and foreign invasions from virtually every direction between 1817 and 1836, Oyo gradually disintegrated. Cities, towns, and villages were attacked, some were destroyed and others were founded. When populated places (referred to hereafter as ``cities'') were involved in conflict, those conquered were routinely captured, enslaved, and transported in slave caravans along existing trading routes to ports on the Atlantic coast as well as inland into internal slave markets, particularly the Sokoto Caliphate, which was founded in 1804 and is located to the northeast of the Oyo region we study in this work. 

Historian and co-author Henry B. Lovejoy \citeyearpar{lovejoy2019} describes how he compiled a list of cities in Oyo involved in conflict from 1817--1836, their corresponding spatial coordinates, start and end dates surrounding conflict at each city, political affiliations of the city, and the primary or secondary sources from which the data derive. A conflict intensity scale is encoded as a categorical variable with two levels: $2$ indicates a city was attacked and $3$ represents a city that was destroyed. Cities could be attacked or destroyed over several years. Cities were often rebuilt in the same or slightly different locations and attacked or destroyed again in subsequent years. New cities were also founded, which is represented with $9$ in the data set. The data are available in our data repository at \url{https://osf.io/h6upw}. As historians continue refining their understanding of conflict in Oyo, additional degrees or levels of conflict could be added to the data set. In Section \ref{SS:krig} we describe how we create maps of conflict density based on this data set.

\subsection{Historical trade routes}
\label{SS:routes}
\begin{figure}[t!]
	\centering
	\includegraphics[width=0.96\linewidth]{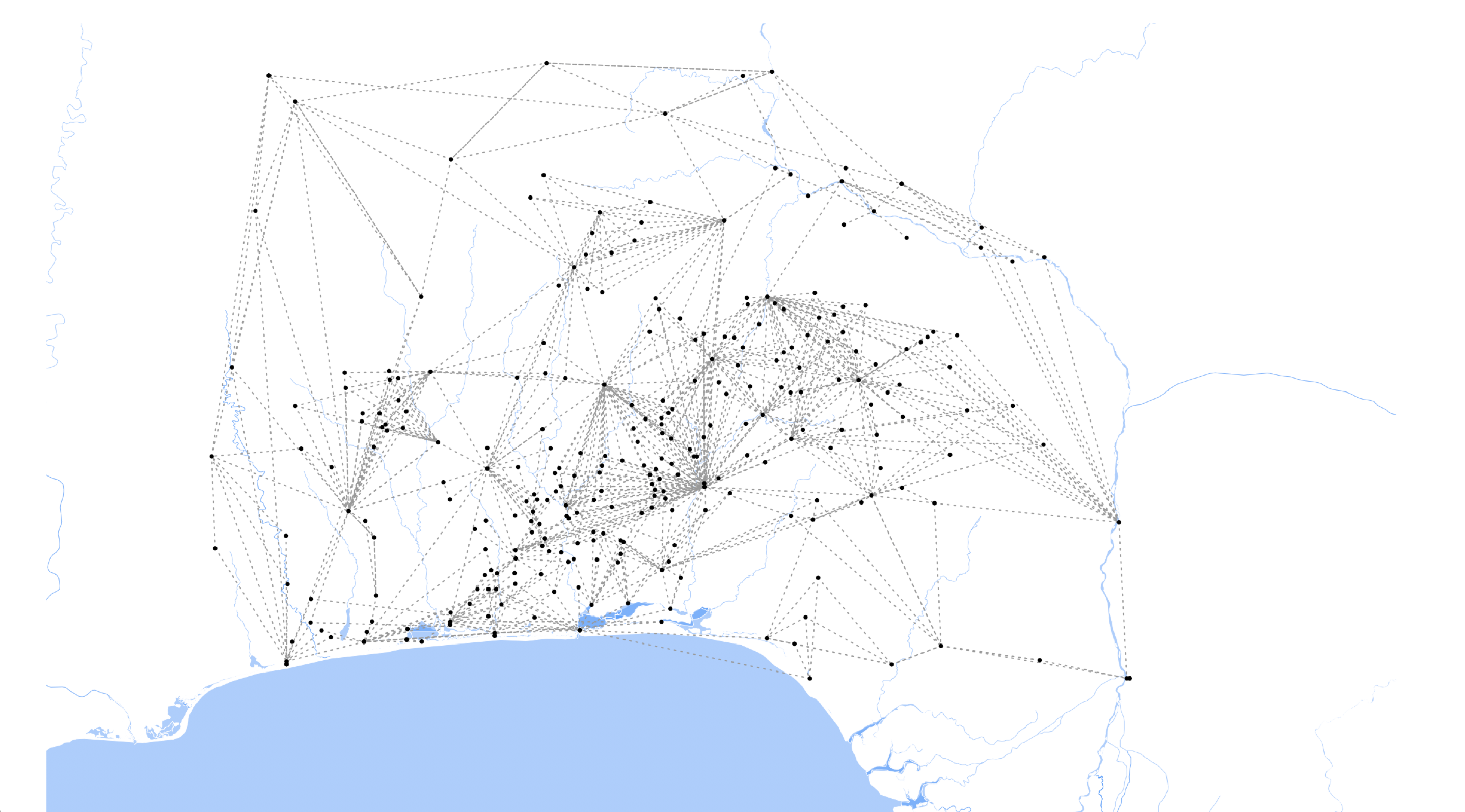}
	\caption{Map of trade in Oyo, 1817--1836, which represents the adjacency matrix for the Markov decision process}
	\label{fig:1824Trademap}
\end{figure}

Lovejoy \citeyearpar{lovejoy2019} also describes how trade routes were constructed from lists and locations of cities in Oyo from 1817--1836. To document the trade network among these cities, an adjacency matrix was compiled to describe the trade routes available as analyzed and debated in the historiography. The trade route adjacency matrix provides historical connections between cities, all of which are connected in some way to potential coastal and inland ports-of-sale. These data are also available at \url{https://osf.io/h6upw}. Figure \ref{fig:1824Trademap} illustrates the possible routes in the trade network. One assumption is that this trade network is constant over time. As Oyo lost control of the region over time, the historical trade routes may have shifted due to arising conflicts and the British Royal Navy's blockades of slave trading ports. This transition is not recorded in these data but could be incorporated into our modelling framework.

We encoded the relationships (edges) between the cities (nodes) of this graph into an adjacency matrix, describing which cities were connected. An adjacency matrix $A$ for a set of $n$ locations (nodes) $s_1, \cdots, s_n$ is of dimension $n \times n$. An entry $A_{ij}$ is equal to 1 if there is a connection starting at $s_i$ and ending at $s_j$, and 0 otherwise. Since trading caravans travelling from $s_i$ to $s_j$ could also travel in the reverse direction, the adjacency matrix describing the historical trade routes is symmetric. We use this adjacency matrix to define the set of available actions in a Markov decision process, described in Section \ref{SS:MDP}.

\subsection{Slave ships and passenger logs}
\label{SS:slaveships}
The \textit{Voyages: The Trans-Atlantic Slave Trade Database} contains data for 243 documented slave ships departing several ports in the Bight of Benin from 1817--1836, and \cite{eltis2020} provide annual estimates of the total number of enslaved people departing this region using these data. Many documented voyages have missing data on the port, year of departure or arrival, or the number of enslaved individuals embarking on ships. Some researchers have applied statistical techniques such as sampling using Markov chain Monte Carlo methods to estimate total embarkations by decade and region of Africa for slave ships with missing data \citep{manning2015,manning2020}. Since these imputations are by decade and region (not year or specific port), we use the estimated annual departures data from Eltis \citeyearpar{eltis2020}.

While specific documentation for who was on board slave ships do not exist for most ships, starting in 1807 the British Royal Navy captured many slave ships, escorted them into bilateral courts of mixed commission in Freetown and Havana, and compiled detailed lists of the enslaved people forced on board. These detailed passenger lists included transliterations of African names which could be interpreted to determine a name's language (whether Yoruba, Fon, Hausa, among others) for use in future analyses surrounding the unique ethnolinguistic composition of individual slave ships, as contemplated in Nwokeji and Eltis \citeyearpar{nwokeji2002}. In the future, linguistic data for transliterated African names might be aligned on an annual basis to the internal zones of conflict and probabilistic origins we are attempting to identify in this paper. Henry B. Lovejoy \citeyearpar{lovejoy2016} has interpreted upwards of 4,000 documented African names from ship registers in Havana, although there is another data set for some 15,000 individuals documented at Freetown, Sierra Leone \citep{d1}.

\section{Models for Migration Due to Conflict}
\label{S:Models}
The models we use to describe historical migration due to conflict are comprised of two parts. First, we create a conflict surface from documented points of conflict and then use this surface as a density function representing points of capture and enslavement. Then, the simulated locations of the captured and enslaved individuals are used as input to a model describing forced migration through Africa. The models used to create the conflict surface and internal migrations are discussed in the following sections.

\subsection{Mapping historical conflict density}
\label{SS:krig}
Building on the historiography, Lovejoy \citeyearpar{lovejoy2019} has recorded major events surrounding the fall of Oyo, such as the kingdom's borders collapsing inwards, Dahomey achieving independence to the west, the founding of the emirate of Ilorin in jihad to the east, and lost territory to a coalition of neighboring Yoruba-speaking kingdoms to the southeast. While the historical borders of the resulting countries are relatively well known, answering the question of from where enslaved people originated requires analyzing the conflicts themselves to determine the regions within the greater Oyo area most impacted by each conflict. Within our region and time period of study, we have geo- and time-referenced data on instances of conflict and the destruction of cities (with examples of protracted warfare recorded as less intense than the complete destruction of cities), as described in Section \ref{S:Data:Conflict}. However, a conflict does not unfold over just the sites of the major battles and events: armies mobilise, advance, raid, retreat, occupy and abandon cities---thereby generating enslaved individuals---throughout contested regions \citep{cameron}. To account for this behavior based on highly uncertain historical data, we created a continuous spatial map of the regions of conflict given the discrete time and place data available.

\subsubsection{A Gaussian process for generating slave capture locations}
The data consist of conflict intensity measures $\boldsymbol{Y}$ observed at 2-D spatial locations $\mathbf s_1, \mathbf s_2, \dots \mathbf s_n$. The statistical process model is

\begin{align*} 
    Y(\mathbf s) &= Z(\mathbf s) + \varepsilon(\mathbf s), 
\end{align*}

where $\varepsilon(\cdot)$ is assumed to be an independent mean zero Gaussian white noise process with nugget variance $\tau^2$, representing measurement error or microscale variation. $Z(\cdot)$ is modeled as a mean zero spatially correlated Gaussian process with Mat\'ern covariance function. In particular, the Mat\'ern covariance function is
\begin{align} \label{eq:Matern}
  \operatorname{Cov}(Z(\mathbf s), Z(\mathbf s')) = k(\mathbf s,\mathbf s') = 
  \sigma^2\frac{2^{1-\nu}}{\Gamma(\nu)}
  \Bigg(\frac{d}{\kappa}\Bigg)^{\nu}
  K_{\nu}\Bigg(\frac{d}{\kappa}\Bigg),
\end{align}
where $d=\|\mathbf s-\mathbf s'\|$ is the Euclidean distance between points, $\sigma^2$ is the marginal variance (or sill), $\kappa>0$ is the spatial range parameter and $\nu>0$ is the smoothness.  $\Gamma$ is the gamma function, and $K_{\nu}$ is the modified Bessel function of the second kind of order $\nu$. The Mat\'ern covariance function is a popular choice in spatial statistics because it is a flexible model with interpretable parameters \citep{stein1999}. 

The Gaussian process considers the data $\boldsymbol{Y}$ to be a single draw from a multivariate normal on $\mathbb{R}^n$.  This yields a log likelihood proportional to

$$\ell(\boldsymbol{Y}) \propto -\log\det\left(\Sigma+\tau^2I\right)-\boldsymbol{Y}^T \left(\Sigma+\tau^2I\right)^{-1} \boldsymbol{Y},$$
where $I$ is an $n\times n$ identity matrix, and $\Sigma$ is given by the Mat\'ern covariance function in Eq. \ref{eq:Matern} with the $(i,j)^{th}$ entry defined by $ \Sigma_{i,j}=k(\mathbf s_i, \mathbf s_j)$ .

The formal kriging estimator fills in a map of the Oyo region with a conflict intensity measure at chosen resolution by taking each desired location $\mathbf s_0$ and computing the estimated conflict intensity $$\hat{\boldsymbol{Y}}(\mathbf s_0)= k(\mathbf s_0,\mathcal{S}) \left(\Sigma+\tau^2I\right)^{-1} \boldsymbol{Y},$$ where  $k(\mathbf s_0, \mathcal{S}) = (k(\mathbf s_0, \mathbf s_1), k(\mathbf s_0, \mathbf s_2), \cdots, k(\mathbf s_0, \mathbf s_n))$ is a $1\times n$ vector with entries defined by the Mat\'ern covariance function evaluated pairwise at $\mathbf s_0$ and every point in $\mathcal{S} = \{ \mathbf s_1, \mathbf s_2, \dots \mathbf s_n \}$.

Instead of a Gaussian process, there are several alternative models which could be used to convert the discrete spatial observations (sites of conflict) into a smooth spatial surface. These include using splines or a kernel density estimator (KDE) to do the smoothing, or treating the discrete observations as a realisation of a point process model. While a spline or a KDE may have the advantage of computational efficiency over a Gaussian process, they may not provide enough flexibility. For example, using the Gaussian process model allows one to adjust the marginal and nugget variances, the spatial range, and the smoothness parameters using the data, as opposed to only a kernel bandwidth parameter in the case of a KDE.

\subsubsection{Estimating kriging model parameters}
To create the conflict intensity surface from discrete conflict data, we identified appropriate covariance parameters with a variogram using the discrete conflict data from the year 1832, which had the highest number of observations of conflict (45) for any single year in the period 1817--1836. The range was chosen to be $\kappa=0.13$ degrees latitude/longitude, the smoothness $\nu=3$, the sill $\sigma^2=0.4$, and the nugget $\tau^2=0.1$. 
While this is very smooth in the context of dense spatial data, our observations are quite sparse, and higher smoothness helps ensure that our maps of conflict intensity can include ridge-like structures between separated conflicts, which mimic shifting borders resulting from the ebb and flow of warfare. Smaller values for the smoothness and range parameters would enforce a more rapid decay to zero away from observations and would push our model closer to one found from kernel density estimation with small bandwidth because it would result in conflicts being modeled as small, radially symmetric, disconnected, additive kernels around the observed locations. 
We did not include a mean part of the statistical model so that the conflict intensity surface decays to zero away from conflict. Thus, these four covariance parameters are all that are required to perform spatial kriging at any desired location.

In the classical kriging sense, this surface exists in the units of the $\boldsymbol{Y}$, which is the intensity of conflict at a city. However, we can also view the resulting surface as an implied density function, where the larger values of the surface near conflict locations represent regions of increased probability of capture and enslavement. By using the kriging estimator to fill in a high-resolution surface over the Oyo region for each year or any set of years under investigation, we can create a probability density function by normalising by the sum total of all predictions on the grid as follows. Formally, we predict $\hat{Y}(\mathbf g_i)$ on a regular rectangular grid $\mathbf g_i$ for $i = 1, \cdots, n_g$, and then normalise these predictions to create a probability density function

\begin{align} \label{eq:Ytilde} 
\Tilde{Y}(\mathbf g_i) = \hat{Y}(\mathbf g_i) / \left( \sum_{i=1}^{n_g} \hat{Y}(\mathbf g_i) \right). 
\end{align}

Since the numerical computation is performed on a discrete regular grid, $\Tilde{Y}$ is a probability mass function. Then, $\Tilde{Y}$ can be used to simulate the origin locations of people enslaved as a result of conflict in the region for any given year or set of years.

\subsection{Modelling the transit of enslaved people}
\label{SS:MDP}
The map in Figure \ref{fig:1824Trademap} displays cities within and around Oyo connected by the most likely trade routes at the time \citep{lovejoy2019}. Such a depiction naturally translates to a graph-based approach to capturing the economics of the region. As described in Section \ref{SS:routes}, the map is derived from an adjacency matrix of valid city-to-city movements. In addition, we classify a handful of cities to be absorbing states where enslaved people historically departed the region externally via Atlantic ports (Lagos, Ouidah, Little Popo, Jakin, Badagry, and Porto Novo) or internally within slave markets (denoted Off Map NE, SE, and NW). Rather than directly assign probabilities to the flow of enslaved people in the region, we apply a finite horizon Markov decision process to model the decisions of slave traders at the time since this family of models can be constructed to emulate sequential decision making \citep{boucherie2017}.

\subsubsection{The historical narrative}
To model the historical transit of enslaved people through the Oyo region, we must consider three historical realities and express them in statistical terms in our transit model. First, while historical trade routes have mostly been reconstructed by historians \citep{lovejoy2019}, the exact routes from points of capture to points of sale, which could be used to inform a Markov decision process, remain largely unknown. Historically, slave routes changed over time, shifting in response to conflict and perhaps differing personal preferences of slave traders. These factors almost certainly resulted in a variety of plausible paths throughout the region. Consequently, our transit model requires sufficient variation to allow for enslaved people captured in similar locations to deviate to different ports of departure simply by chance.

Second, much of the slave trade had to pass through pre-determined cities for the collection of taxes or tributes, as well as protection. Additionally, areas of conflict could ensnare slave traders causing them to be captured and enslaved. As a result, slave traders were incentivised to avoid areas of conflict perhaps to seek protection from enemies or other hostile situations \citep{ajayi1967,curtin1967,lloyd1967}. Statistically, this means that we require a model for the transit of enslaved people to be able to downweight probabilities of transit based on conflict intensities as well as distance travelled.

Finally, the traffic of enslaved individuals at the ports of sale shifted over time, with a generally eastward preference among slave traders over time from Ouidah to Lagos. This shift was largely in response to British anti-slavery blockades becoming more prominent in front of Ouidah \citep{a4}. Therefore, to align our model with the historical narrative, our method must allow for some sale locations to be preferred to others \textit{a priori}, whether that preference is informed by volume of trade, the price of slaves, or the gradual west-to-east blockade of West African slave ports by the British Royal Navy's African squadron during suppression efforts of the transatlantic slave trade. 

\subsubsection{A Markov decision process}
A Markov decision process (MDP) describes the partially deterministic and partially stochastic movement of an agent through a network in discrete time. The agent's actions at each state are chosen based on the rewards and costs associated with reaching future states in the network, and the actual event that takes place can be probabilistic. Formally, an MDP consists of a 5-tuple $(S, A, P_a, R_a, \gamma)$. $S$ is a finite set of states in a network, often spatially located. $A$ is the set the actions an agent can take from any given state $s \in S$. $S$ and $A$ can also be thought of as the nodes and edges in a network, respectively. In our case, $S_y={\{s_1,\dots, s_{n_y}\}}$ is the set of locations of the $n_y$ cities in the trade network in year $y$. $A_y$ is an adjacency matrix of the trade network in year $y$ such that $a_{ij}=1$ if and only if cities with locations $s_i$ and $s_j$ are directly connected, i.e., a slave trader could travel from $s_i$ to $s_j$ without passing through a third city $s_k$. The trade networks defined by $S_y$ and $A_y$ are actually constant from year to year in our study period, but could change as more historical data become available. This set of historical trade nodes $S$ and edges $A$ for Oyo from 1817--1836 is described in Section \ref{SS:routes}.

For an action $a \in A$ taken in state $s$ at time $t$, we define the probability $P(s_{t+1}' \, | \, s_{t}, a)$ of reaching state $s'$ for all states in $S$. This probability has a Bernoulli distribution with the two outcomes of moving to the new state or remaining in place. We place a small probability of remaining in state $s$, i.e., $P(s_{t+1} \, | \, s_{t}, a)=0.02$.  In practice, this approach allows for some possibility of a slave caravan staying in the same city for a number of iterations, which opens avenues for non-stationarity in time that we do not further explore.

Rather than \textit{a priori} preferring one immediate move through the network over another, we allowed the paths taken in the MDP to be determined by optimising the reward/cost incurred after moving from $s$ to $s'$ via action $a$, which we write as $R(s_{t+1}' \, | \, s_{t}, a)$. The overall reward $R = R_T - R_M$ includes negative values ($-R_M$) that represent cost of movement through conflict and positive values ($R_T$) that correspond to transporting an enslaved person to a point-of-sale absorbing state. Movement along each edge in the network incurred a cost proportional to $D*(1+\lambda C)$, where $D$ is the length of that edge, $C$ is the maximum of the conflict kriging estimate $\Tilde{Y}$ along that edge, and $\lambda$ is a scaling factor for conflict. Formally, 

\begin{align*} 
R_M = \sum_{t=0}^{n-1}{R(s_{t+1}' \, | \, s_{t}, a)} = \sum_{t=0}^{n-1}\frac{D_{ss'}  (1 + \lambda C_{ss'})}{C_{max}} \times \mathbbm{1}_{t+1}(s \xrightarrow{} s') ,
\end{align*}

where $\mathbbm{1}_{t+1}(s \xrightarrow{} s')$ is the indicator function of the agent moving from state $s$ to $s'$ at time $t+1$ and $C_{max}$ is the maximum value of $\Tilde{Y}$ over the entire grid $\{\mathbf g_i\}_{i=1}^{n_g}$.

Positive rewards $R_T$ can be incurred in the Markov decision process by reaching an absorbing state. An absorbing state is defined in our case as a city in which the only possible movement to be taken is to itself. This outcome is encoded as a row of the adjacency matrix with a 1 on the diagonal and 0's elsewhere. Movement into these states represented a sale, gaining a terminal reward. Absorbing states hold the positive values of the rewards $R_T$. 
We set terminal rewards $R_T$ for the set of $n_A$ absorbing states to be normally distributed with a common mean and a variance $\sigma_R^2$ reflecting the slave traders' (random) personal preferences and imperfect information, i.e., $R_T \sim N(\bm{\mu}_R, \sigma_R^2 \boldsymbol{I})$, where $\bm \mu_R$ is a vector length $n_A$ and $I$ is an $n_A \times n_A$ identity matrix.

The MDP solves the problem of finding an optimal policy $\pi^{*}(s)$ whose value specifies the action $a$ to take by the agent at state $s$. In our case, the function $\pi$ is found by maximising the expected total rewards: 
$$E^{\pi}\left[\sum^{\infty}_{t=0} {\gamma^t R_{a_t} (s_t, s_{t+1})}\right].$$
The discount factor $\gamma \in [0, 1]$ allows the rewards incurred in later time steps to be downweighted. We fixed the discount factor at $\gamma=1$ because we wanted the cost of movement to increase relative to the distance travelled and conflict traversed rather than by the number of steps taken or cities visited.

Many algorithms have been developed to solve this optimization problem, e.g., linear or dynamic programming. We optimised the total reward criterion using the policy iteration algorithm implemented in the \texttt{R} package \texttt{MDPtoolbox} \citep{chades2017} to find the optimal policy $\pi^{*}(s)$. The policy fixes the action to be taken at any step in the MDP, which reduces the probability matrix $P$ into a Markov transition matrix showing the probabilities of moving from state $s$ to $s'$. In our case, $P(s_{t+1}' \, | \, s_{t}, a)=0.98$.

Given a draw from the distribution of $R_T$, the end result of the policy iteration algorithm is a ``best route" for a slave trader to reach a point-of-sale from any given origin location, which is determined by the trader's personal preferences and/or their imperfect information about the actual terminal sale rewards as well as the cost of moving through conflict. Due to the random deviations we encoded in the reward $R_T$, this route may be potentially different for each slave caravan, even for those originating in the same location.

\subsubsection{Aligning the MDP with the historical narrative}
Modelling the transit of enslaved people with this MDP formulation allows for considerable flexibility in meeting our criteria for a transit model that aligns with the historical narrative to allow for variability in paths taken from point of capture to point-of-sale, avoiding transit through conflict areas, and non-uniform rewards at points of sale.

Randomness in absorbing rewards $R_T$ can account for both individual slave trader preferences and the broader temporal shift from the western port of Ouidah to the eastern port of Lagos. Setting all point-of-sale rewards to be equal asks the question, ``What is the least resistance route to \textit{any} point-of-sale," whereas varying the reward vector allows for individual slave traders to balance preferred or higher revenue sale locations with the implied costs of a longer journey or a journey through regions of conflict.

The cost-benefits formulation of MDP reward maximization can be adjusted to downweight specific movements. In our case, we explicitly make movement through regions of conflict less desirable. More generally, this formulation could be expanded to include disincentives to cross borders, 
venture through certain terrains, or deviate from historical trade routes. Each of these possibilities are in addition to the distance-based cost terms with which we initialise the model.

Once the conflict intensity surface $\hat{Y}$ is created via kriging, $\Tilde{Y}$ (see Eq. \ref{eq:Ytilde}) can be repeatedly sampled to generate simulated people enslaved due to conflict. Captured individuals can be passed to the MDP with either varying or identical reward vectors $R_T$, modelling the movement of the individual through the trade network to an absorbing city. This approach allows  us to create a large sample of enslaved individuals and their eventual points-of-sale. These considerations can be used to describe the ultimate goal of this study: from which locations did enslaved people originate inland before departing specific ports? 

We further discuss the use of unequal rewards to add an appropriate amount of stochasticity to these simulation results in Sections \ref{SS:Tuning} and \ref{SS:FinalMaps}.

\subsubsection{MDP parameter tuning}
\label{SS:Tuning}
Due to the lack of historical information on the price of slaves as a function of point-of-sale, we choose as a baseline an MDP with equal expected rewards $\bm{\mu_R} = 10\cdot\bm{1}_{n_A}$ with variance $\sigma_R^2 = 0.1$ for each absorbing state, where $\bm{1}_{n_A}$ is a unit vector of length $n_A$.

As described in the previous section, movement along each edge incurred a cost proportional to $D*(1+\lambda C)$, where $D$ is the length of that edge and $C$ is the maximum of the conflict kriging estimate along that edge. This approach allows conflict in an area to affect both the origin locations of enslaved individuals and the transition chains to ports of sale.

Varying the conflict scaling factor $\lambda$ results in conflict contributing to the decision making in the MDP, resulting in different totals of enslaved individuals arriving at terminal nodes. To estimate an optimal value of $\lambda$ based on currently available data, we ran the model combining simulating individuals from the conflict density surface and using them as input to the MDP until they reached an absorbing state. To get the most accurate value of $\lambda$, we aggregated the conflict and ship total data from all years. The ship totals data let us verify the output for our model against historical data. The ship total data is sparse for some years, which is why we chose to aggregate the ship total data and conflict data over all years. We let $\lambda$ vary in the optimisation in order to minimise the difference between the expected totals at each port $E_i$ calculated by our model and the observed totals at each port $O_i$ given in the ship totals data set. We seek to minimise 

$$\sum_{i=1}^{n_P} \frac{(E_i - O_i)^2}{E_i},$$

where $n_P$ is the number of absorbing states. This has the form of a $\chi^2_{n_P-1}$ statistic. The results of this optimization yield an optimal value of $\lambda = 1.55$. This value is then used in the model on an annual basis to produce conditional probability maps for each year, described in the next section.

\subsection{Mapping conditional probabilities of origin}
\label{SS:FinalMaps}
\subsubsection{Annual large-scale simulation}  
To gain origin and departure information from the models in Sections \ref{SS:krig} and \ref{SS:MDP}, we generate many enslavements of individuals from the conflict density function. Then, for each such enslaved individual, we generate a random terminal reward vector $R_T$ from a distribution specifying the end reward for selling that individual at any given absorbing state in the network. We fit an MDP for each enslaved individual and reward vector pair, resulting in an optimal policy and eventual path of motion for each enslaved individual. We perform this simulative approach on an annual basis and we use the conflict data specific to each year from 1817--1836 to create the conflict density function and then generate 1,000 samples per year used as input to the MDP model.


\subsubsection{Kernel density smoothing for maps}
To integrate these simulations---at this point a large collection of origin points encoded by their point-of-departure---into more cleanly interpreted spatial maps, we use a simple kernel density estimator, which creates a small, radially decaying kernel function at each simulated slave capture location leaving from a specific port. The aggregation of each kernel function for every enslaved individual leaving a given port results in a heat map for enslavement origin given the port of departure. As historians often know the port of departure of enslaved people, we can use the repeated samples of simulated data to estimate probabilities of origins for the enslaved individuals who left a given port in a given year.

Formally, the kernel density estimate takes a radially symmetric function $K(r)$ and estimates the regional heat map $\hat{f}$ via the weighted sum

$$\hat{f}(x)=\frac{1}{nh}\sum_{i=1}^n K\left(\frac{|x-x_i|}{h}\right), $$

where $x_1, x_2, \dots x_n$ are the $n$ capture locations for the enslaved people departing from the port in question. We choose the multivariate normal as the radial function $K$, as is often convention \citep{terrell1992}. In general, a kernel density estimate requires only tuning one parameter: the bandwidth $h$ that determines the distance/width of the kernel function centered on each simulated slave capture location. The \texttt{R} function \texttt{kde2d} in package \texttt{MASS} implements the multivariate normal kernel density as its default and is employed here \citep{venables2002}. While a kernel density function can be sensitive to the number $n$ of points employed, our simulation-based model allows us to simulate any arbitrary number of spatial samples and construct the resulting kernel density estimator to desired precision. In the web application described in Section \ref{S:App} we allow $h$ to vary from $0.25$ to $6$ degrees latitude/longitude for a sample of 1,000 simulated enslaved individuals per year and find that this range provides visually acceptable maps. A larger simulated sample would in turn allow for smaller bandwidths.

The sum of kernels corresponding to all simulated individuals approximates the conflict density function as more individuals are simulated. The density can be decomposed based on which point(s) of sale the simulated individual ended in via the MDP. To generate a proper probability density function that integrates to 1, we normalise the sum of the kernels for any subset of the points of sale. Numerically, for each point(s) of sale, the KDE is predicted on a regular grid, with predictions in grid cells that lie in the ocean set to 0. Then the predicted values are normalised by the sum of all KDE predictions on the entire grid. This outcome results in a probability estimate, summing to 1, conditional on the individuals ending up in the selected absorbing state(s)

\subsection{Model summary}
Our method for using historical data to estimate probabilities of origins of enslaved people during the collapse of Oyo from 1817--1836 works in four steps.
\begin{enumerate}
    \item Using space-time locations for historical conflict, for each year, create a map estimating the conflict density $\hat{Y}$ that represents the shifting borders of the wars involved. Normalise $\hat{Y}$ into a proper density function $\Tilde{Y}$, then sample enslavement locations from this conflict density function.
    \item Create a trade map based on historical roads, cities, and common trading routes at the time. Specify the cities known to be historical points-of-sale on this map.
    \item For each sample generated in step 1, create an MDP using the adjacency matrix in step 2 by pairing it with a randomised terminal reward vector $R_T$. Record each enslaved individual's point of capture, route, and point of departure. The MDP includes at least two flexible parameters: the cost-of-movement through conflict scaling factor $\lambda$ and the variance of the terminal rewards $\sigma_R^2$ for each sale location.
    \item For each port of departure, create a heat map of enslavement origins via kernel density estimation.
\end{enumerate}
    
\section{Results}
\label{S:Results}
Results from the modelling approaches developed in this work can inform the possible origins and paths of enslaved individuals, as well as relative numbers of individuals arriving in each port of departure. 

\begin{figure}[t!]
	\centering
	\includegraphics[width=\linewidth]{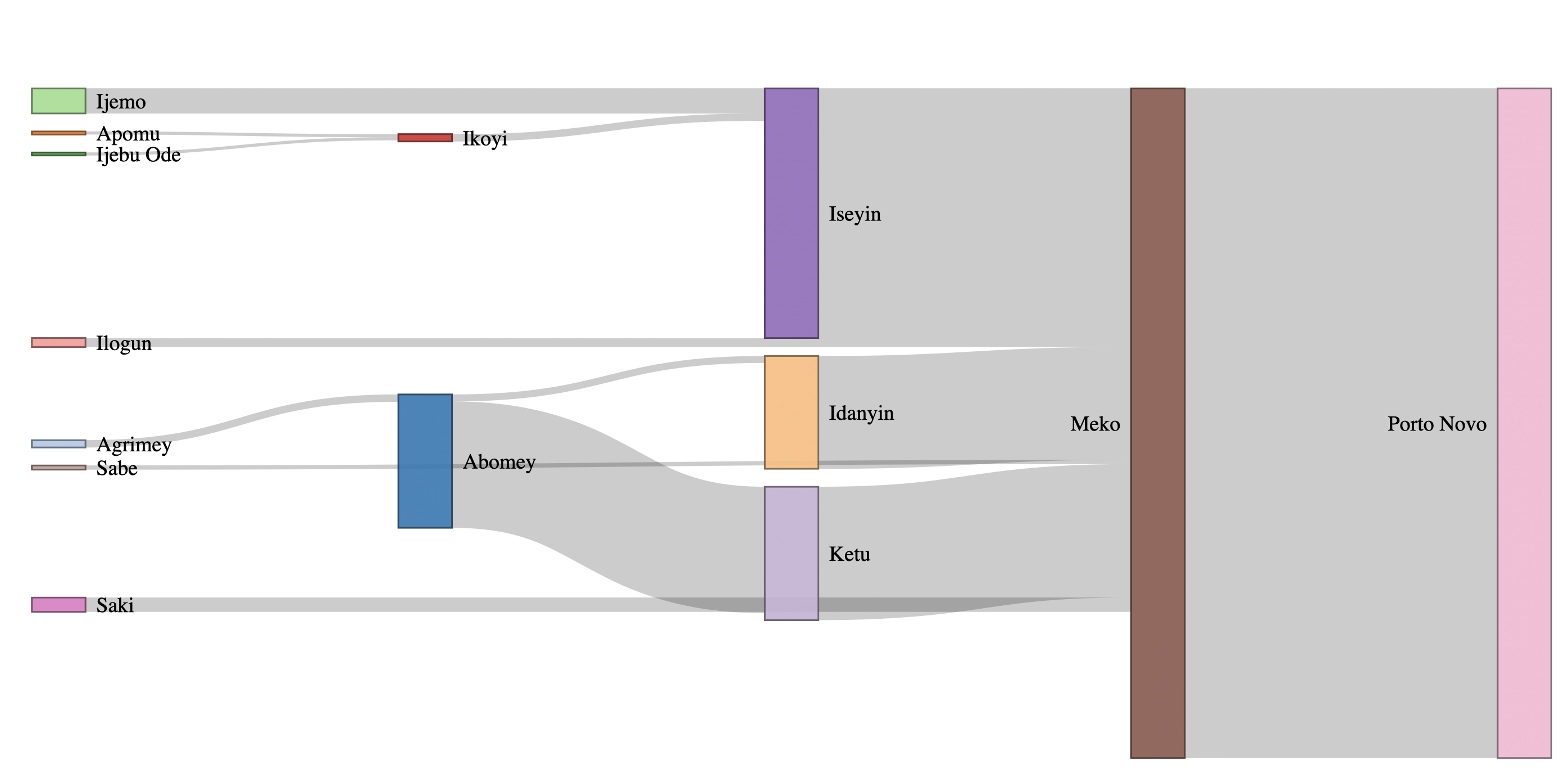}
	\caption{Sankey diagram showing the paths of all individuals arriving in Porto Novo over all years 1817--1836.}
	\label{fig:Sankey}
\end{figure}

First, we present a Sankey diagram in Figure \ref{fig:Sankey}, which displays the possible origins and paths of enslaved individuals who departed from the absorbing state Porto Novo based on our model aggregated over all years. The width of the line connecting cities indicates how many individuals traveled between these two cities. We do not display absolute numbers, and the size of the lines are to be interpreted relative to each other because the figure is based on how many individual simulations were produced in our model. It is likely that those who came through Porto Novo came through major centers of trade such as Abomey, Iseyin, Idanyin, and Ketu, while a much smaller number of individuals originated from places such as Ijebu Ode or Ijemo. 

Another way to visualise the results of the model are through the conditional probability plots described in Section \ref{SS:FinalMaps}. For each year from 1817--1836 and for each point-of-sale, we estimated the conditional probabilities of origin for an enslaved person leaving that port. In this section we present eight conditional probability maps: the first set of four illustrate the conditional probabilities of origin from four different ports from our model for the year 1824, and the second quartet illustrate the conditional probabilities of origin for an enslaved individual leaving any coastal port in the years 1828, 1829, 1831, and 1832. 

\begin{figure}[t!]
	\centering
	\includegraphics[width=1\linewidth]{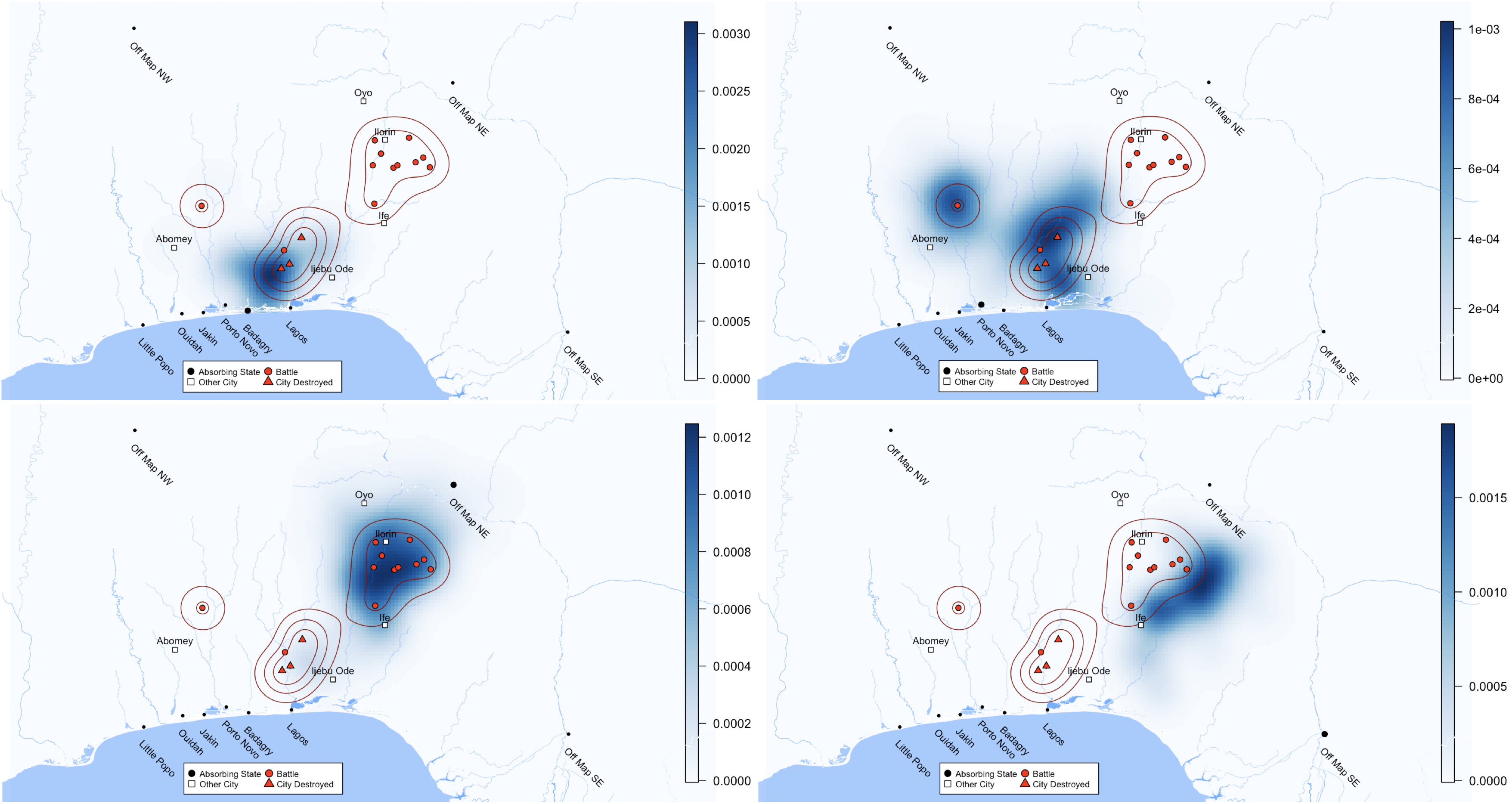}
	\caption{Conditional probabilities for 1824. The top left panel shows the probability of origin given the enslaved person departed Badagry. The top right shows the same for Porto Novo; bottom left is given they departed Off Map NE into the Sokoto Caliphate; bottom right for departed Off Map SE. The contours of conflict are displayed along with battles (circles) and cities destroyed (triangles).}
	\label{fig:1824montage}
\end{figure}

In comparison to the Sankey diagram above, Figure \ref{fig:1824montage} shows the conditional probabilities of the origins of enslaved people departing from Badagry, Porto Novo, Off Map NE, and Off Map SE panelled clockwise starting in the top left. Off Map NE can be thought of as an enslaved individual leaving Oyo and entering into the Sokoto Caliphate. 

These four maps demonstrate the likely origins of enslaved people involved in Oyo's collapse on a port-by-port basis. By providing these visualisations from our model, it is immediately possible for historians of the African diaspora to begin to assess from which inland cities people originated before boarding slave ships for the Americas. The red triangles represent cities being destroyed, and the red circles represent cities involved in protracted warfare. To summarise Figure \ref{fig:1824montage}, an enslaved person captured during Oyo's collapse in the year 1824 leaving through coastal ports most likely came from  conflicts in the South, whereas an individual leaving via the Off Map NE or Off Map SE absorbing states most probably came from conflicts in the North.

This model output can lead to more specific historical commentary. The top left panel shows that people leaving Badagry almost certainly arrived due to conflict at cities in the Egba, Awori, Egbado, and Anago regions. The top right panel illustrates the arrival of people embroiled in conflict due to an Ijebu, Ife, and Ijesha alliance along the Osun river. Much of this conflict would have also involved southern Oyo migrations due to the expansion of Ilorin in jihad as an emirate within the Sokoto Caliphate. The bottom left panel shows how people engaged in conflict also stemming from Ilorin were likely absorbed into the slave trade of the Sokoto Caliphate in the northeast. The final panel illustrates the likely areas for enslaved people being take off map to the southeast whereby small numbers of people likely went into the kingdom of Benin or into trade networks along the Niger River. 

\begin{figure}[t!]
	\centering
	\includegraphics[width=1\linewidth]{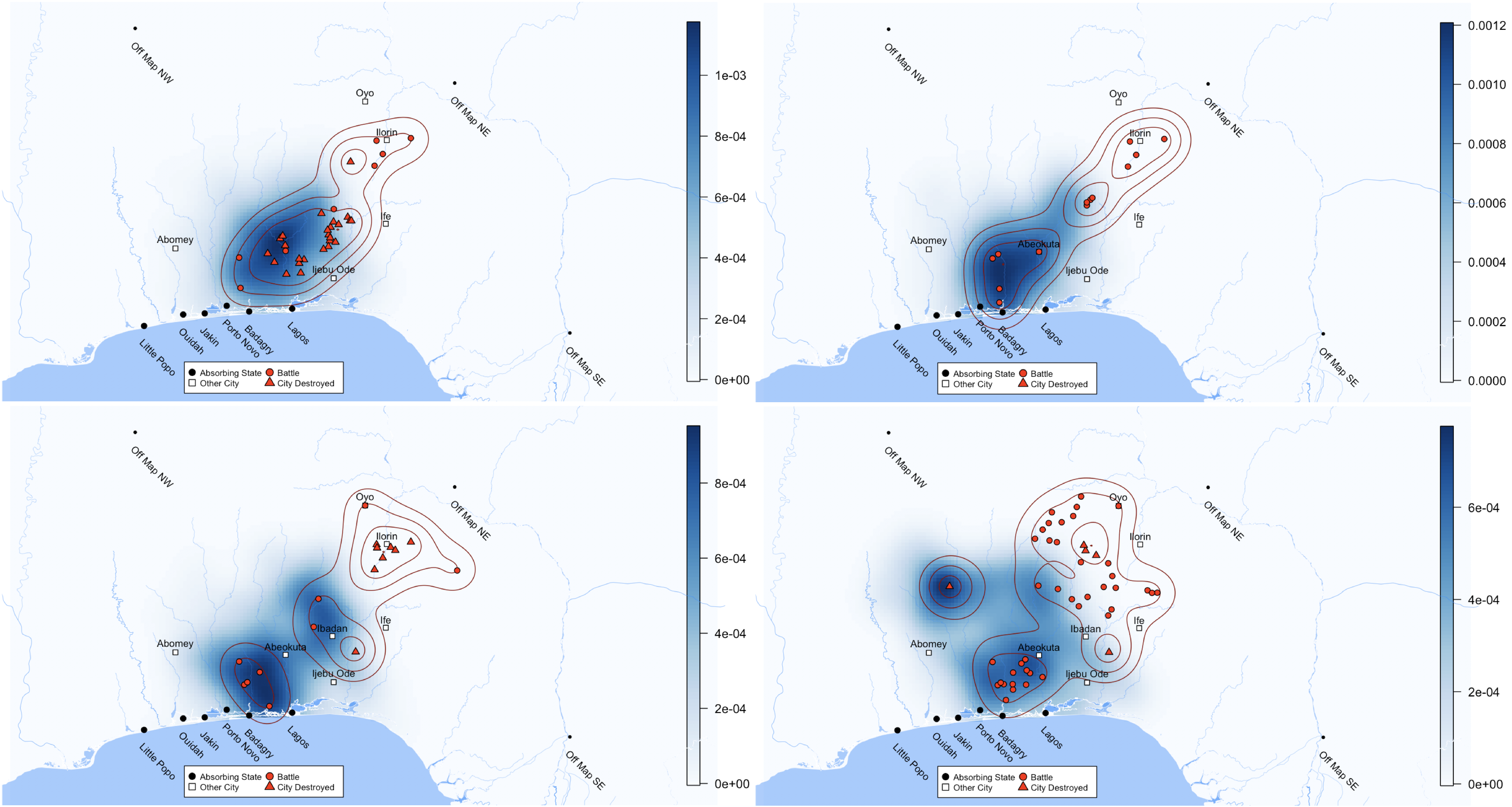}
	\caption{Conditional probabilities of origin of enslaved people departing Oyo from any coastal port into the transatlantic slave trade. The four panels represent conditional probabilities of origins for 1828, 1829, 1831, and 1832 clockwise starting in the top left}
	\label{fig:4yearcoast}
\end{figure}

Figure \ref{fig:1824montage} is a selection of results, and the model and application can produce similar results for every port for every year between 1817--1836. Figure \ref{fig:4yearcoast} illustrates another sample of results from our model, showing the conditional probabilities of origins for enslaved people departing Oyo from any coastal port, including Porto Novo, Lagos, Ouidah, Jakin, or Badagry, for the years 1828, 1829, 1831, and 1832.

These four maps demonstrate different years and shifting zones of conflict due to the ongoing pressure on Oyo from all fronts. This figure highlights one of the primary features of our model: the probable origins of enslaved persons leaving the coastal ports (or any absorbing state) change throughout the years based on the areas embroiled in historical conflict. 

Historical events can be connected to the shifting zones of origin probability. In 1828, represented by the map on the top left, the Ijebu Ife alliance continued to attack the Egba and Egbado regions, while Ilorin forced Oyo refugees further south destroying a series of cities. Meanwhile, Dahomey continued to push eastward into the Egbado, Anago, and Awori regions. The following year, in the map displayed on the top right, Ilorin slave raiding was widespread throughout Oyo territory and the city of Ibadan was founded as a refugee center in the south. Initially, Ife armies dominated the new city, but as Oyo and Owu refugees arrived, they became strong enough to expel the Ife chiefs. In 1831 (the map on the bottom left), several key Oyo cities were abandoned in part due to the drought of 1830 and the consolidation of another refugee center of predominately Egba and Owu at Abeokuta. Meanwhile, Ibadan---as a new city-state---began to wage war, although Abeokuta repulsed those attacks. Dahomey in the west continued its systematic campaign in the Egbado and Awori regions. In 1832 (the map on the bottom right), most of the Oyo cities came under attack by Ilorin in the north. In the south, the Owiwi war involved an unsuccessful attempt by Ijebu and support from Ibadan to disrupt trade routes to Abeokuta. Meanwhile, Dahomey attacked from the west and destroyed the Mahi city of Kpaloko in the north. In terms of enslaved people entering Atlantic networks, Eltis (2020) estimates that nearly 7,800 enslaved people departed these coastal ports for the Americas and Sierra Leone in 1832. Their likely origins were mostly from the Egba, Egbado, Awori, and Anago regions, and from the southern fringes of Oyo around Ibadan. Others would have included Ijebu and Ife, among other groups from the forest, who were captured in conflict by their enemies. An assessment of those migrations would likely result in similar totals of people being transported to the north off map into Sokoto.  

\subsection{Interactive Web Application}
\label{S:App}
To make this research more widely available to a general audience, we created an interactive web application using the \texttt{Shiny} package in the \texttt{R} programming language \citep{chang2019}. The user can select a year and one or more points-of-sale, and the application generates and displays a conditional probability map showing the most likely region of capture based on our model. The app can also display the yearly conflict data as discrete points or a contour plot of the conflict density surface. Furthermore, the trade network informing the MDP and annual approximate state borders \citep{lovejoy2019} can be overlaid. 

The data that is visualised in the application come from running our model for each year from 1817--1836. For each year, we generate 1,000 capture locations and record their spatial coordinates, the initial location in the trade network, and the point-of-sale. For any set of ports we produce an annual conditional probability surface using the methods of Section 3. For each year, we save the conditional probability surface, the conflict point data, the kriging conflict density surface, the trade network, and the state border shapefiles, which are all the data sets required to host the app. The state borders are the same those visualised at Yoruba Diaspora (\url{http://slaveryimages.org/}), which is a cartographically based interactive digital archive currently under development by Vachan Daffedar Aswathanarayana and Brumfield Labs \citep{d2}, and our app can be compared to the more detailed maps at the previous link.

Our web application is easy to use and freely hosted at \url{http://dev.regid.ca/YD}.
It enables a general audience to interactively explore the history of the West African slave trade by visualising the data and models used in this paper. Note that we do not claim that these maps display the historical truth, but rather the results from a model which provide an approximation of the truth. As historians learn more about the West African slave trade, improved data can lead to more accurate results from our model.

\section{Discussion}
\label{S:Discussion}

In this work, we developed a method for using historical data describing conflicts and trade routes in Oyo from 1817--1836 to estimate the uncertain origins of people enslaved in this region during this time period. We used kriging to produce annual conflict density surfaces, from which we simulated the capture of individuals. We modeled the transport of enslaved people to ports of sale using an MDP. Then given an enslaved individual's location of departure, we created maps of that person's probability of origin. Our \texttt{Shiny} web application can produce such maps for any of eight departure points in the region during any of the years between 1817 and 1836. 

\subsection{Contributions to historical literature}

The first innovation of this work was to create continuous maps of conflict via kriging in pre-colonial Africa during the era of the slave trade, which may have an immediate and long-lasting impact of the field of African and African diaspora history. By organising geo-political data on a temporal axis and aligning it with slave voyage data, historians can quickly access annual maps to visualise slave ship departures juxtaposed with the occurrence of conflict. Since slavery usually begins with some act of violence, the conflict provides a clearer and easier means in which to present these data. Over the long term, historians could add and refine data to generate a broader understanding of how inland origins of enslaved people changed alongside the constant ebb and flow of conflict within Africa.

The second innovation of this work was to combine conflict data with historical trade routes to estimate likely origins of enslaved people leaving the Oyo region from 1817---1836. For historians engaged in digital humanities projects, creating such annual maps of the probable origins of enslaved individuals can spur future research and innovation. As an example, our work may help to classify origins of individuals in growing DNA and genealogical databases, such as 23andMe.com, Ancestry.com, or FamilySearch.org, which appear only able to provide broad regional origins \citep{ancestry2014, genetic2015, 23andme}. If a person today knows that they descend from someone who was transported on a specific ship leaving the Oyo region, our methods and maps can provide a much richer understanding of their ancestral origins.

Our results based on conflicts in Oyo indicate that many individuals were enslaved due to conflicts around Ilorin and then likely transported north into the Sokoto Caliphate. Paul E. Lovejoy has argued that ``the internal trade of [enslaved individuals in] West Africa appears to have been on a scale that was comparable to that of the transatlantic slave trade," which suggests our model could be reverse engineered to re-evaluate the scale of demographic change within Africa during the precolonial period \cite[p.~104]{c2}. If our model is able to withstand critique for its treatment of the transport of enslaved individuals from Oyo to coastal areas, it could be applied to new data in the Sokoto Caliphate to help provide better estimates for the number of people involved in historical migrations into the Central Sudan.

Estimating the probabilities of origin for the rest of pre-colonial Africa will require much more support in terms of extracting and collating geo-political data, which must be accomplished region-by-region, city-by-city, and year-over-year. Input from African scholars and universities will be required, especially as new archaeological discoveries emerge. Ancient cities are being unearthed, many of which no longer have a place name or reflect the present-day geography. Archaeologist Akinwumi Ogundiran has, for example, identified remains of cities that are currently not located on modern-day or historical secondary source maps, but could have been places destroyed in the conflicts surrounding Oyo's collapse \citep{akinwumi2, akinwumi3, akinwumi4, shadow2007}. Planning for this long process of data collection and assimilation has begun, whereby a team of historians have re-regionalised Africa with a vocabulary that expressly implements a more neutral terminology to avoid terms associated with European slave traders, colonial states, and modern-day countries, which often confuse the representation of inland Africa before 1900 \citep{vocabulary2020}.

Mapping conflict and origin locations of migrants in larger regions over longer periods in pre-colonial Africa will undoubtedly provide monumental results capable of making deeper connections to the complex heritage of African ancestry around the world by linking together existing and emerging open source data sets, specifically those that have a biographic focus on enslaved Africans and their descendants. Due to increased activity in digital humanities practices, large data sets are in the ongoing process of being curated and expanded; and their themes are wide-ranging with projects centered on abolitionism, baptisms, self-liberating slaves, slave owners, slave narratives, and other land or maritime migrations beyond the Atlantic World \citep{d1, d2, d3, d5, d6, d7, d8, d9, d10}. Connecting these data will provide richer analysis for the intersection of people and historical sources at different moments in time. Filling the void of information surrounding pre-colonial Africa's shifting geo-political landscapes and migrations will revolutionise how African history is taught and understood, while providing a more empowered voice to the enslaved person's experience.

\subsection{Potential extensions}
Future extensions of our methods could include estimating a population density surface for our region of study, and---instead of drawing a captured individual from the conflict density map---we could simulate enslavements from the product of the population surface and the conflict density surface, thereby reflecting the reality that more people would be captured from densely populated areas of conflict than from sparsely populated conflict areas. Such historical population density estimates do not currently exist, but historians could create such estimates in the future, similar to those in Manning \citeyearpar{Manning2010}. 

One of the advantages of our model is its flexibility to incorporate better sources of data. For example, as more slave ship logs are transcribed and the names categorised into linguistic groups with regional origins, more data will become available to tune the model's parameters. Additionally, genetic databases and genealogical tracking have become much more powerful in recent years \citep{larmuseau2015, mathias2016,primativo2017,gouveia2020}, and we look forward to an increase in the availability of such data. In particular, if descendants of passengers of any known ships were to compare their genome to the current genetic mapping of the Bight of Benin inland areas, we could improve the model validation. Our model could also be adapted to account for the time variance in the process of transporting captured individuals to ports-of-sale and the lack of precision in recorded conflict dates. One option would be adding positive spatio-temporal correlation from one year's conflict map to the next. Another option would be to add a time delay to each step in the MDP, allowing for recalculations of optimal routes as the conflicts shift each year.

A final source of tuning and validation would be to fit similar models to similar historical situations with better availability of data. Mapping continuous conflict borders from discrete city observations could be done for nearly any conventional war fought in the 18\textsuperscript{th} or 19\textsuperscript{th} centuries. Forced transit situations in the Holocaust did not originate from conventional battles as in Oyo, but have considerably better data due to its relative recency, and could be used to better tune the decision process and exit location models.

\subsection{Additional applications}
Beyond the specific application of understanding the origins of enslaved individuals involuntarily involved in the transatlantic slave trade, our methods would translate well to other migrations of marginalised groups including the depopulation of indigenous groups in the Americas, the Jewish diaspora, contemporary refugees, as well as others associated with genocide and slavery. Specifically, our methods could help map the origins of refugees displaced during the contemporary Syrian Civil War, during which cities are known to have been attacked or destroyed, causing the migration of hundreds of thousands of individuals \citep{williams2020}. Given the known networks of roads and first-hand accounts of paths taken by Syrian migrants, an MDP could be created to model the migration of these displaced individuals into neighboring countries. Given a specific refugee camp or border crossing, a map of probabilities of the origins of these migrants could be created to provide an estimation of from where in Syria these migrants/refugees originated. Such a model could be validated by known demographics and places of origin for a sample of refugees \citep{ozden2013, adali2020}. Similarly, Olsson and Siba \citeyearpar{olsson2013} have compiled a list of 530 towns attacked in the Sudanese region of Darfur, and our methods could be used to better understand the resulting migrations of Darfuris. Ambrosius and Leblang \citeyearpar{ambrosius2019} argue that gang violence in El Salvador increases demand to emigrate to the United States. Our methods could be helpful to generate evidence for or against this claim.

\section{Conclusion}
\label{S:Conclusion}
This paper describes a method for using historical data on conflicts and trade routes in Oyo from 1817--1836 to create a map of conflict using kriging, simulating the capture of enslaved people in regions of conflict, simulating the transit of these people to ports of sale, and then estimating their probabilities of origin given that they left the region from a specific port. We believe that this method could be generally applicable to other situations of forced migration due to documented conflict. We look forward to the continued development of statistical models to provide insight for historians by creating maps in the presence of uncertainty, modelling historical migrations, and estimating the probabilities of origins of these migrants, and making those tools available in a non-proprietary form to the public and academic researchers.

While this paper is limited to one quadrant of Africa in a twenty-one year period, we view these results as an iterative process, whereby we are working with a limited set of geo-political and conflict data that only paints a partial picture for the broader region. Adding more data and context will invariably change results presented herein. Regardless, the approaches, maps, and preliminary results related to Oyo developed herein tie together and produce visualizations of the extensive secondary historical literature for this region. Scaling this project to encompass a broader geographic and temporal framework will require historians, computer scientists, statisticians, and geneticists working in tandem. Knowing the highest degrees of probabilities for when and where people most likely came from inland in all of Africa will have major ramifications for the history of the Atlantic world, whereby the ocean \emph{connects}, rather than disconnects, Africa, the Americas, and Europe. In order to achieve this end, more inland conflict data will need to be collected for other African regions across lengthier periods of time, and then aligned with slave voyage data. The results would undoubtedly improve our understanding of African ancestry on all sides of the Atlantic, and eventually the Indian Ocean world too. We look forward to larger, interdisciplinary collaborations that will uncover more insights associated with one of the world's largest, forced transoceanic migrations in human history.

\section*{Acknowledgements}

We would like to acknowledge the support of the Andrew W. Mellon Foundation, as well as the Graduate School and the Center to Advance Research and Teaching in the Social Sciences at the University of Colorado Boulder. We would also like to thank Vachan Daffedar Aswathanarayana for helping us mount the R Shiny App onto servers.  

\bibliographystyle{rss}
\bibliography{Oyo.JRSSC.main}

\end{document}